# Shear transformation zone analysis of shear reversal during granular flow

Michael L. Falk[1,2*], Masahiro Toiya[3], Wolfgang Losert[3*]

[1]Department of Materials Science and Engineering, Department of Mechanical Engineering, and Department of Physics and Astronomy, Johns Hopkins University, Baltimore, MD 21218, USA

[2] The Kavli Institute for Theoretical Physics, University of California, Santa Barbara, CA 93106, USA

[3]Department of Physics, University of Maryland, College Park, Maryland 20742, USA

*email: mfalk@jhu.edu; wlosert@umd.edu

Experimental measurements of the onset of granular flow are directly compared to predictions of the "shear transformation zone" (STZ) theory of amorphous plasticity. The STZ equations make it possible, on a coarse grained level, to incorporate the anisotropy of the particle contact network and its dynamic evolution including changes in the contact network direction under shear reversal. The STZ theory and experiment show qualitative agreement in the transient and steady state shear forces and velocity profiles during the start of shear and shear reversal.

## I. Introduction

Quantifying flow in granular systems has long been an area of interest because of the importance of granular media in industrial applications and earth systems. However, the nature of granular flow differs in many ways from both solid deformation and the rheology of liquids [1]. Most striking is the emergence of a yield stress below which flow becomes unsustainable, and the emergence of mesoscale features such as force chains [2]. Understanding these phenomena is complicated by the fact that the large particle sizes make the effect of thermal fluctuations negligible. The result is that the disordered yet highly non-ergodic nature of the granular packing presents a significant challenge theoretically, as it lacks both the properties of long-range order and self-averaging [3].

Similar issues arise in the low temperature behavior of amorphous solid materials. Materials like metallic glasses exhibit highly disordered structures that remain far from equilibrium [4-6], although in practical applications, particularly near the glass transition temperature, thermal effects do play a significant role in these materials [7-8]. Here we test the applicability of the shear transformation zone (STZ) theory [9], which was developed in the context of such amorphous solids and recently reviewed by Falk and Langer [10], to the flow in a granular material. To do so we consider the low temperature limit of this theory, and use it to describe the flow in a Taylor-Couette cell during the onset of shear and shear reversal.

We have chosen to focus on shear reversal as a key process that involves the granular microstructure. While much work on granular materials focuses on steady flowing states or jammed states, many natural and industrial flows are short lived. The transition from a jammed to a flowing state has proven very difficult to incorporate within a consistent theory. Reversal of the direction of shear, in particular, involves a change in the principal stress axes by 90 degrees. Since the force network and contact network have been shown to be highly aligned with the principal stress axis [11], shear reversal requires the buildup of a new force bearing contact network in a direction orthogonal to the existing network. Since the load bearing contact network is a crucial feature of steady granular flows, shear reversal offers the opportunity to investigate how the steady contact network builds up with shear strain at the start of granular flow. The STZ theory provides a framework to include such anisotropies. We find that the STZ theory allows for a description of the full transition from jammed state to steady flow, and is able to predict history dependent non-steady processes in granular flows.

## II. Experimental Methodology and Results

Our experimental apparatus consists of two concentric cylinders with a moving inner cylinder and a fixed outer cylinder. The bottom plate is attached to the outer cylinder. The gap between the cylinders is filled with spherical glass beads of diameter 1mm or 2mm. The inner cylinder radius, $r_{in}$, is 102mm and the gap between the cylinders, $\Delta r$, is 44mm. A layer of glass beads of the same kind that fills the gap is glued to both cylinder surfaces.

The top surface of the granular material is imaged from above at up to 500 frames/s. A torque sensor monitors the torque exerted on the inner cylinder. In a typical run the inner cylinder is rotated at 6-12 mHz, corresponding to a tangential velocity of 4-8 mm/sec. Particles on the top surface appear as bright spots in the images. A particle tracking algorithm in IDL identifies these bright spots, and tracks them from frame to frame.

The inner cylinder is rotated at a slow steady speed to prepare the particles in a reasonably consistent initial arrangement prior to the start of each run. We find that the start of shear flow is not strongly dependent on shear rate at the speeds we investigated from 0.3 mHz to 0.5 Hz. This is consistent with other findings for slow steady shear [12-15]. The inner cylinder is stopped after initialization. Shear flow is initiated again by rotating the inner cylinder in the opposite direction.

The primary observations from our shear reversal experiments have been reported [14] and are illustrated by the data in Figure 1. Here six lines show the angular velocity measured in six concentric rings during the experiment described above. When sheared in the initial direction the system reaches steady state during which three of the six rings do not experience any appreciable flow. If the driving is discontinued and then reapplied, the system reaches the same steady state immediately. However if the driving is stopped and then applied in the reverse direction, a transient sets in during which flow is evident both in the regions that were flowing and in the previously jammed regions. Average flowing velocities in all regions are faster initially and drop off with roughly the same time scale in all regions.

**III. Review of STZ Theory in the Athermal Regime**

In order to acquaint the reader with the physical derivation of the STZ theory we will review the basic premises here with an emphasis on elucidating the origin of the tensor internal state variables that arise in the theory. The basic assumptions used in constructing the STZ equations are the following postulates.

1. In the amorphous solid at any given time there exist a dilute concentration of local regions that are susceptible to shear; we refer to as shear transformation zones (STZs).
2. These regions are not susceptible to any arbitrary shear, but only to shear along a particular slip orientation.
3. Such slip events are, at least temporarily and approximately, reversible so that the STZs act as microscopic two-state systems.
4. These regions have a characteristic size.

The assumption that STZs are two-state systems is necessary for the master equation to exhibit distinct flowing and jammed states and dynamic cross-over between these states at a well defined flow stress.

Once these assumptions are made then the implication is that we can express the local structure of the amorphous material in terms of a local distribution of STZs. This orientation dependent distribution varies between material points, and it's angular dependence must be symmetric about the origin. Additionally, the assumption that an STZ transition causes the zone to flip a perpendicular orientation implies that there is a special relationship between any orientation and the corresponding perpendicular orientations.

Rather than specifying the distribution of STZs at every point in space, it is preferable to consider only the first few moments of this distribution. To this end we can calculate the first two non-zero moments of the distribution as a scalar parameter, $\Lambda$, that corresponds to the average number of STZs of all orientations, and a traceless symmetric tensor parameter **m**. The distribution of STZs in the direction $\Omega$ is then proportional to

$$P(\vec{r},\Omega) \propto \Lambda(\vec{r})\left[1+\mathbf{m}(\vec{r}):\mathbf{d}(\Omega)\right], \tag{1}$$

where **d** is a unit director tensor in the direction $\Omega$, which in 3D can be expressed as

$$\mathbf{d}(\Omega) = \sqrt{\tfrac{3}{2}}\left[e_\Omega \otimes e_\Omega - \tfrac{1}{3}\mathbf{I}\right], \tag{2}$$

where $e_\Omega$ is a unit vector in the $\Omega$ direction, and **I** is the identity matrix.

At this point it is simpler to derive our equations of motion in terms of the number of STZs along the direction and perpendicular to the direction of the applied deviatoric stress **s**, the traceless part of the full stress tensor $\boldsymbol{\sigma}$. We will use $N_+$ to refer to the number of STZs oriented along the same orientation as **s**, and $N_-$ to refer to the number with a perpendicular orientation. Doing so provides more physical transparency in the derivation without sacrificing generality since we can readily recover the full tensor form of the equations of motion. To do this we define the number of zones aligned with the applied stress as

$$N_+ = N\Lambda\left[1+\mathbf{m}:\mathbf{d}\left(\tfrac{\mathbf{s}}{s}\right)\right],$$
$$N_- = N\Lambda\left[1+\mathbf{m}:\mathbf{d}\left(-\tfrac{\mathbf{s}}{s}\right)\right] = N\Lambda\left[1-\mathbf{m}:\mathbf{d}\left(\tfrac{\mathbf{s}}{s}\right)\right]. \tag{3}$$

Here $N$ is simply the number of atoms in the system and $s$ is the magnitude of **s**.

We can now express the rate of plastic strain as proportional to the rate at which these two species of zones undergo transitions using a kinetic rate equation [16]. The rate of plastic deformation $D^{pl}$ depends on these two species and an as yet unspecified rate function $R$.

$$D^{pl} = \frac{\varepsilon_0}{N}\left[R(s)N_- - R(-s)N_+\right] \quad (4)$$

Here $\varepsilon_0$ is the strain per shear transformation. In addition, we need to write a master equation for the populations of STZs. We assume from the outset that we may do this in a mean-field sense. This equation is written in the form

$$\dot{N}_\pm = \left[R(\pm s)N_\mp - R(\mp s)N_\pm\right] + \Gamma(s, N_\pm)\left(\tfrac{1}{2}N_{eq} - N_\pm\right) \quad (5)$$

where the term in brackets provides for the inter-conversion of STZs oriented along the direction of applied stress and perpendicular to this direction. The second term on the right hand side provides, in a mean-field sense, for the non-linear terms that must arise when STZ transitions result in the creation and annihilation of neighboring STZs. It is assumed that the same mixing rate, $\Gamma$, is responsible for both processes. A rewriting of Eqs. 4 and 5 in terms of $\Lambda$ and **m** results in equations of motion for the tensor rate of plastic deformation, scalar state variable $\Lambda$ and the tensor state variable **m** of the form

$$\mathbf{D}^{pl} = \varepsilon_0 C(\mathbf{s})\Lambda\left[\mathcal{T}(\mathbf{s})\frac{\mathbf{s}}{s} - \mathbf{m}\right], \quad (6)$$

$$\dot{\Lambda} = \Gamma(\mathbf{s}, \Lambda, \mathbf{m})(\Lambda_{eq} - \Lambda), \quad (7)$$

and

$$\dot{\mathbf{m}} = \frac{2\mathbf{D}^{pl}}{\varepsilon_0 \Lambda} - \Gamma(\mathbf{s}, \Lambda, \mathbf{m}) \frac{\Lambda_{eq}}{\Lambda} \mathbf{m}. \qquad (8)$$

Here the scalar functions $C$ and $\mathcal{T}$ are composed from the original rate function $R$, as

$$C(\mathbf{s}) = \tfrac{1}{2}\left[R(\mathbf{s}) + R(-\mathbf{s})\right] \qquad (9)$$

and

$$\mathcal{T}(\mathbf{s}) = \frac{R(\mathbf{s}) - R(-\mathbf{s})}{R(\mathbf{s}) + R(-\mathbf{s})}. \qquad (10)$$

At this point there are three unresolved aspects of the constitutive model. The first is the value of $\Lambda_{eq}$, the steady state density of STZs. For the purposes of this analysis we will assume that due to the nature in which the granular packing is initially created, the STZ number density has already reached its steady state value. Often, as discussed in Ref. [10], the evolution of the STZ density can be quite important for understanding the dynamics of amorphous solids, but given that this particular granular system is driven for a long time in a slow steady state prior to shear reversal, this is an appropriate choice for the set of experimental observations we seek to understand here. This assumption has the result that that Eq. 7, and the $\Lambda$ dynamics become uninteresting, and Eq. 8 is somewhat simplified.

The second unresolved aspect is the functional form of $\Gamma$. This has been treated in some detail in Ref. [10], and here it will suffice to observe that in the granular system we expect mixing only arises from mechanically induced transitions, not from thermally activated events. For this reason there can be only one time scale in the problem, and that must be proportional to the rate of plastic work being done on the system. For this reason we can write

$$\Gamma(\mathbf{s},\Lambda,\mathbf{m}) = \frac{\mathbf{s}:\mathbf{D}^{pl}}{\varepsilon_0 \Lambda_{eq} s_f} \tag{11}$$

The third unresolved aspect is the functional form of the rate function $R$, which provides information about the stress-dependent rate of STZ transitions. For granular materials transitions should only be possible, or at least should be extremely rare, unless the stress is in a direction that would drive the transition. Assuming that only transitions in the driving direction occur, regardless of the detailed functional form of $R$, $\mathcal{T}(s)$ is unity.

The $C$ function is the most phenomenological part of the STZ equations and the most likely part to be dependent on the particular material system in question. In order to work with a set of equations with as few parameters as possible we choose a physically motivated one-parameter set of functions originally proposed in [17]. This set of functions has the form:

$$C(\tilde{s};\zeta) = \frac{|\tilde{s}|^{\zeta+1}}{\sqrt{2\pi\zeta}} \exp\left[-\zeta(|\tilde{s}|-1)\right] + \left(|\tilde{s}|-1-\zeta^{-1}\right) P(\zeta+1,\zeta|\tilde{s}|) \tag{12}$$

Here $\tilde{s} = s/s_f$, and $P$ represents the regularized incomplete gamma function. The parameter $\zeta$, controls how sharply transitions are suppressed as the stress drops below the flow stress.

Given these observations Eq. (6)-(8) simplify to the following two equations:

$$\mathbf{D}^{pl} = \varepsilon_0 C\left(\frac{s}{s_f};\zeta\right) \Lambda_{eq} \left[\frac{\mathbf{s}}{s} - \mathbf{m}\right], \tag{13}$$

$$\dot{\mathbf{m}} = \frac{2}{\varepsilon_0 \Lambda_{eq}} \left(\mathbf{D}^{pl} - \frac{\mathbf{s}:\mathbf{D}^{pl}}{2s_f}\mathbf{m}\right). \tag{14}$$

Here Eq. (13) provides an equation for the strain rate as a function of the current stress and structural state of the material, **m**. Eq. (14) is the equation for the evolution of the structural state. There are two ways that the system may be in steady state. If $\mathbf{m} = \frac{\mathbf{s}}{s}$, then Eq. (13) requires that $\mathbf{D}^{pl}$ is identically zero and as a result the right hand side of Eq. (14) must also be zero, and the structural state, **m**, does not evolve in time. Alternately, it could be the case that $\mathbf{D}^{pl}$ is non-zero, and $\mathbf{m} = \frac{2 s_f \mathbf{D}^{pl}}{\mathbf{s}:\mathbf{D}^{pl}}$. In this case the structural state also does not evolve in time, but the material is flowing. These two steady states coexist when the magnitude of the stress is equal to the flow stress $s_f$, with the non-flowing steady state stable at low stress and the flowing steady state stable at high stress.

**IV. STZ Analysis of Experimental System**

To analyze the experimental results we begin with Eqs. (13) and (14) and we assume that only the $r\theta$ components of the stress and internal structural parameters are relevant to the analysis. We ignore finite strain effects that have been shown to be negligible in similar analyses [18]. Standard elastic assumptions lead to a straightforward solution for the stress as a function of radius.

$$\tilde{s} \cdot s_f = \sigma_{r\theta} = \frac{\sigma_{in} r_{in}}{r} \tag{15}$$

Here $\sigma_{in}$ is the stress on the inner wall. Elasticity also provides an equation that relates the stress to the motion of the inner wall of the experimental apparatus, as denoted by the angular displacement $\Theta$, in the form

$$\frac{\sigma_{in}}{s_f} = \frac{\mu r_{out}}{2s_f(r_{out}-r_{in})}\left[\Theta - 2\int_{r_{in}}^{r_{out}} dr \frac{\varepsilon_{r\theta}(r,\Theta)}{r}\right]. \tag{16}$$

Here μ is the shear modulus of the granular material, and $\varepsilon$ denotes the plastic strain, which varies with radius and evolves as the inner wall is rotated.

According to our theory, as presented above, the deformation rate in the $r\theta$ direction, $D_{r\theta}^{pl}$, is a function of the shear stress in this direction, $s_{r\theta}$, and an internal state variable $m_{r\theta}$ that tells us something about the bias in the orientation of the STZs within the granular compact, essentially an induced plastic anisotropy. Since we specialize to shear in the $r\theta$ direction and all other components are irrelevant, we drop the $r\theta$ subscripts in the equations to follow. This leaves us with the following equations for our rate of plastic strain:

$$\dot{\varepsilon}_{r\theta} = D^{pl} = \varepsilon_0 \Lambda_{eq} C(\tilde{s};\zeta)\left[\text{sign}(\tilde{s}) - m\right], \tag{17}$$

$$\dot{m} = \frac{2D^{pl}}{\varepsilon_0 \Lambda_{eq}}(1-\tilde{s}m). \tag{18}$$

In addition, we acknowledge that in the case of granular flow there is no natural internal time scale independent of the driving rate. Therefore, we will consider the evolution of the system as a function of the inner cylinder angular displacement, Θ. Derivatives of the form $df/d\Theta$ will be denoted $f'$. Our assumption will be that the rate of STZ activation in the granular material is directly proportional to the absolute value of the rate of the rotation of the inner cylinder. With these choices the equations reduce to

$$\varepsilon'_{r\theta}(\tilde{s},m) = \gamma \lambda C(\tilde{s})\left[\text{sign}(\tilde{s}) - m\right]\text{sign}\left(\frac{d\Theta}{dt}\right), \tag{19}$$

and

$$m(\tilde{s},m) = \frac{\varepsilon'_{r\theta}(\tilde{s},m)}{\lambda}(1-\tilde{s}m). \tag{20}$$

Here $\lambda$ is a dimensionless parameter related to the change in strain per unit change in the $m$ state variable. As such, $\lambda \ln 2$ is the strain required to destroy the maximum possible STZ orientational bias, $m=\pm 1$, by shearing at a stress just above $s_f$. $\gamma$ is a scale parameter that relates the inner cylinder angular displacement rate to the rate of STZ activation.

Reviewing the above equations we must now specify a small number of parameters. The parameter $\mu/s_f$, the shear modulus of the granular compact in units of the flow stress, is chosen to be 100, a typical estimate for the yield strength of solids. The STZ dynamics require us to specify three additional parameters, $\lambda$ and $\gamma$, which describe the strain and rate scales of the plastic dynamics and $\zeta$, which gives a measure of how far below the yield stress plastic dynamics persist. In addition we must specify the initial conditions of the parameter $m$ as a function of radial position.

The plastic parameters were extracted by considering the steady state behavior of the system. By applying Eqns. (19)-(20) to the case where stress is held constant independent of time we can obtain an analytic solution for the strain as a function of time. From this result we can extract expressions for the strain and strain rate in the infinite time limit for the case where $\tilde{s}$ is less than and greater than unity respectively. When $\tilde{s}<1$ the strain increases with applied stress as

$$\varepsilon_\infty = \frac{\lambda}{\tilde{s}} \ln\left(\frac{1-\tilde{s}m_0}{1-\tilde{s}}\right), \tag{21}$$

while if $\tilde{s}>1$ the strain rate increases with applied stress as

$$\varepsilon'_\infty = \lambda \gamma C(\tilde{s}; \zeta) \frac{\tilde{s}-1}{\tilde{s}}. \tag{22}$$

To extract approximate values of the parameters we first calculate steady state strains and strain rates from the experimental data as a function of radius. Then examining this data we estimate that at the current driving rate $\tilde{s} = 1$ at a location $r_f = \sigma_{in} r_{in} / s_f$. This allows us to calculate the value of $\tilde{s} = r_f/r$ as a function of the radial position. In our case $r_f$ is approximately 116 mm in the experiment with 1mm particles and 127 mm in the experiment with 2mm particles. Next we can estimate $\lambda$ by assuming that for our reverse strain experiment $m_0 = -1$. We can fit the values for the final strain outside $r_f$ to the functional form $\varepsilon = \lambda (r/r_f) \ln[(r+ r_f)/(r- r_f)]$ and obtain best-fit values for $\lambda$ of 4% and 0.5% for the 1mm and 2mm particles respectively. We note that these numbers are reasonable orders of magnitude for the strains required to reorient the local STZ distribution. Then for any given value of $\zeta$ we can estimate the value of $\gamma$ by fitting equation (22). Further refinements are made to this estimate using the full numerical solution. We find that a $\zeta$ value of 100 and a $\gamma$ value of 18 000 gives a reasonable fit for both experiments.

In order to show a detailed comparison between the theory and the experiment we compare the displacement of the granular material measured at various radii in the Taylor-Couette cell as a function of inner cylinder displacement. Fig. 2 presents the displacement in each concentric ring subsequent to shear reversal versus the inner cylinder displacement. Fig. 2(a) shows the data for the 1mm particle size and 2(b) shows the data for the 2mm particle size. Blue points represent experimental data, while the red lines represent the solutions of the equations given above.

We have also compared the predictions of the above equations to the force applied to the inner cylinder during the experiments. Fig. 3 shows the results of an experiment prior to which steady state had been reached in the negative direction. First the experiment was restarted in that same direction. The experiment was then stopped and restarted in the positive direction, and finally the inner cylinder rotation was stopped and again restarted in the positive direction. The x-axis in Fig. 3 represents time multiplied by the magnitude of the imposed inner cylinder rotation rate during steady-state flow, ω, times the inner cylinder radius. The blue points represent the experimental data, while the red lines represent the theoretical predictions.

It appears that the STZ equations capture the most salient aspects of the deformation subsequent to stress reversal. In particular, they are able to simultaneously account for the fact that the regions close to the outer wall exhibit jamming after a finite amount of deformation, while the regions close to the inner wall continue to flow. The localization of strain observed here appears to arise due to the inhomogeneity of the stress field, which is highest at the inner cylinder, and the fact that the granular material exhibits a yield stress below which steady-state flow can not be maintained. This behavior differs from shear bands observed in more homogeneous loading environments where shear banding has been observed to arise due to a mechanical softening instability [19]. The STZ equations also capture the observation that the forces measured immediately following shear reversal are lower than those observed during restarting rotation in the same direction, as shown in Fig. 3. This arises due to the reorientation of the STZs by 90° under the influence of the altered stress, as evidenced in the theory by the steady evolution of *m* to a value of opposite sign.

Fig. 4 illustrates how such STZ behavior may correspond to the dynamics in the force chain picture. Just like STZs, force chains aligned to cause jamming in one direction do not give rise to jamming under the reverse driving. Therefore time is required to re-form a force chain network subsequent to reversal [18, 20, 21]. Recent experiments in two dimensions have demonstrated how sensitively the force chain network structure depends on the prior stresses and strains [22]. In essence the STZs could be interpreted to represent collections of particles engaged in force chains along particular orientations.

There is an additional conclusion we can draw from the fact that we must vary $\lambda$ (and to a lesser degree $s_y$) to model systems with particles of different sizes. This implies that there exists a system size dependence which is not unexpected since STZs are multi-particle events. The STZ length scale that is typically observed to be 2-3 particle diameters in simulations [9] and in foams [23], but has not been systematically studied in granular solids, and may be larger.

In summary, our intuition from the force chain picture regarding the reversal of granular flow appears to be consistent with experimental observations, and be captured well by the STZ theory. This first version of a granular STZ model indicates that STZ theory can provide a continuum mechanics formalism built upon microscopic insights that has predictive power regarding important macroscale behavior such as the onset of granular flow subsequent to shear reversal.

MLF acknowledges support of this work by the NSF under award DMR0808704 and, in part under PHY0551164. WL thanks Ricardo Pizzarro for assistance in the shear force experiments and NSF for support under award CTS0457431 and CTS0625890

Figure Captions

**Figure 1**  Angular velocity measured in six concentric rings during granular flow in a Taylor-Couette cell. Driving stops and restarts in the same direction, and is then reversed.

**Figure 2**  Comparison between experimental data (points) and the STZ theory (lines) for the radial displacement of concentric rings in the Taylor-Couette cell.  Displacements subsequent to shear reversal are shown for (a) 1mm and (b) 2mm particles.

**Figure 3**  The normalized driving force in the experiment (points) compared to theoretical predictions (lines).  This system of 1mm particles was prepared by shearing in the negative direction.  The system is restarted in the negative direction, stopped, restarted in the positive direction, stopped, and restarted in the positive direction again.

**Figure 4**  The force chain picture and the STZ picture provide different, but compatible representations of the dynamics of jammed regions subsequent to shear reversal.  In the force chain picture chains that constitute the backbone of the jammed system are broken and reformed.  In the STZ picture shearable regions, called STZs, flip until those aligned to promote shear in the flowing direction are exhausted. These STZs may correspond to the connective links in the force chains.

Figure 1

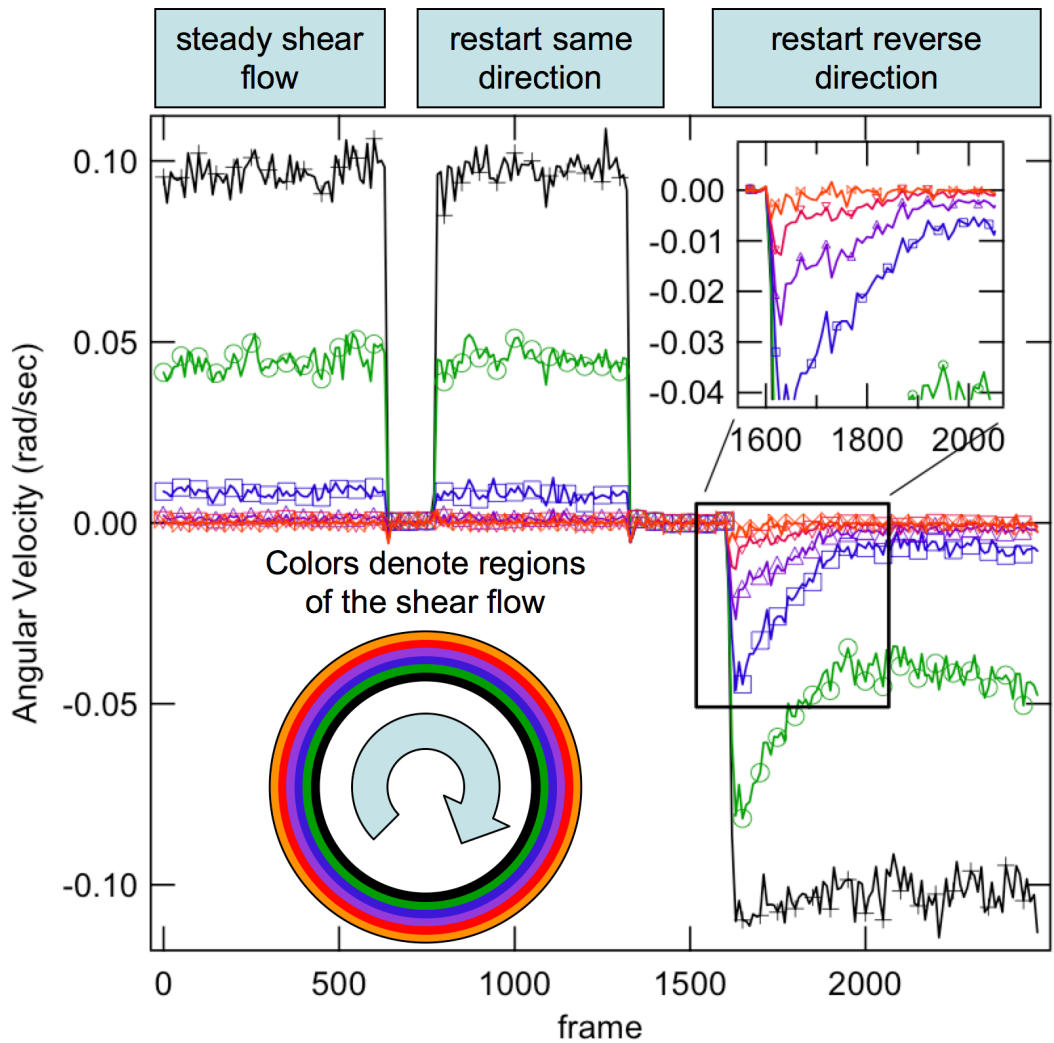

Figure 2

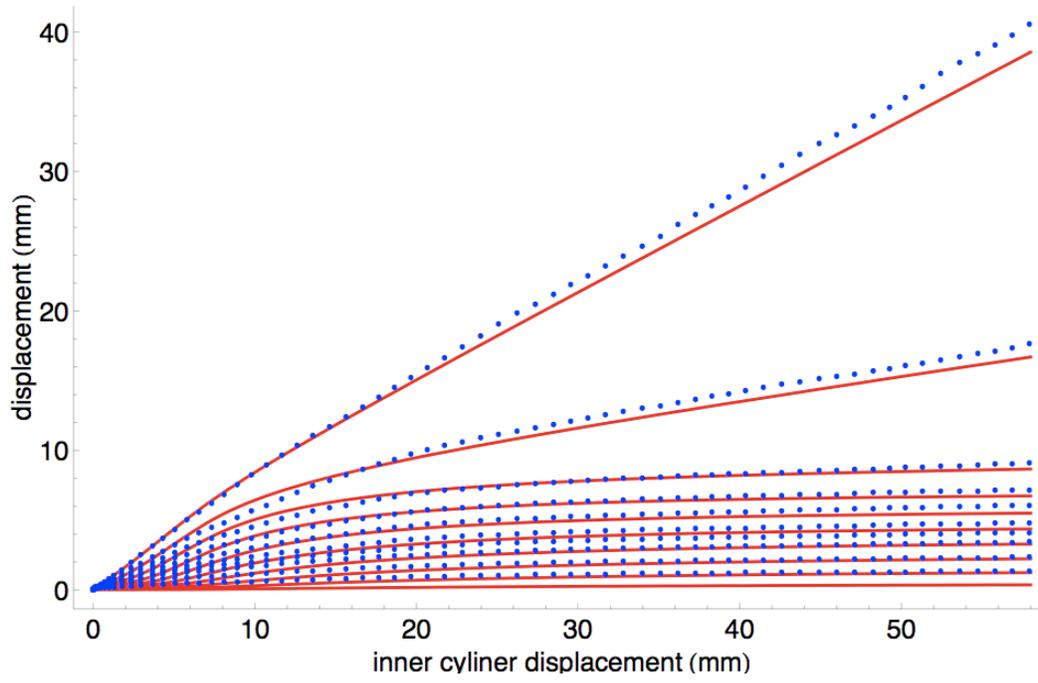

(a)

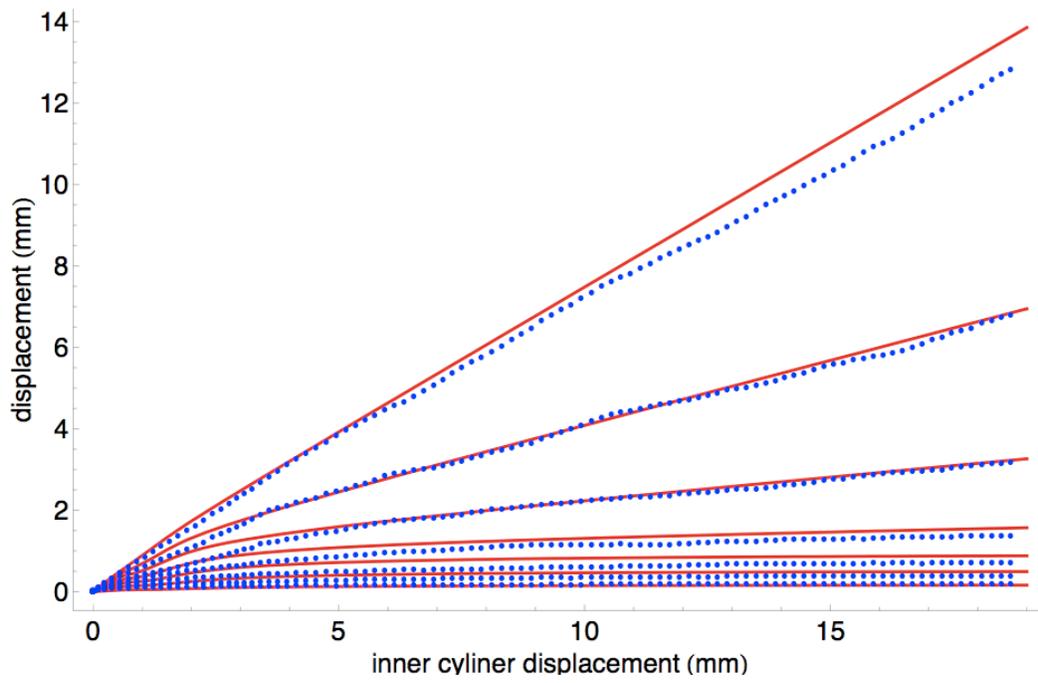

(b)

Figure 3

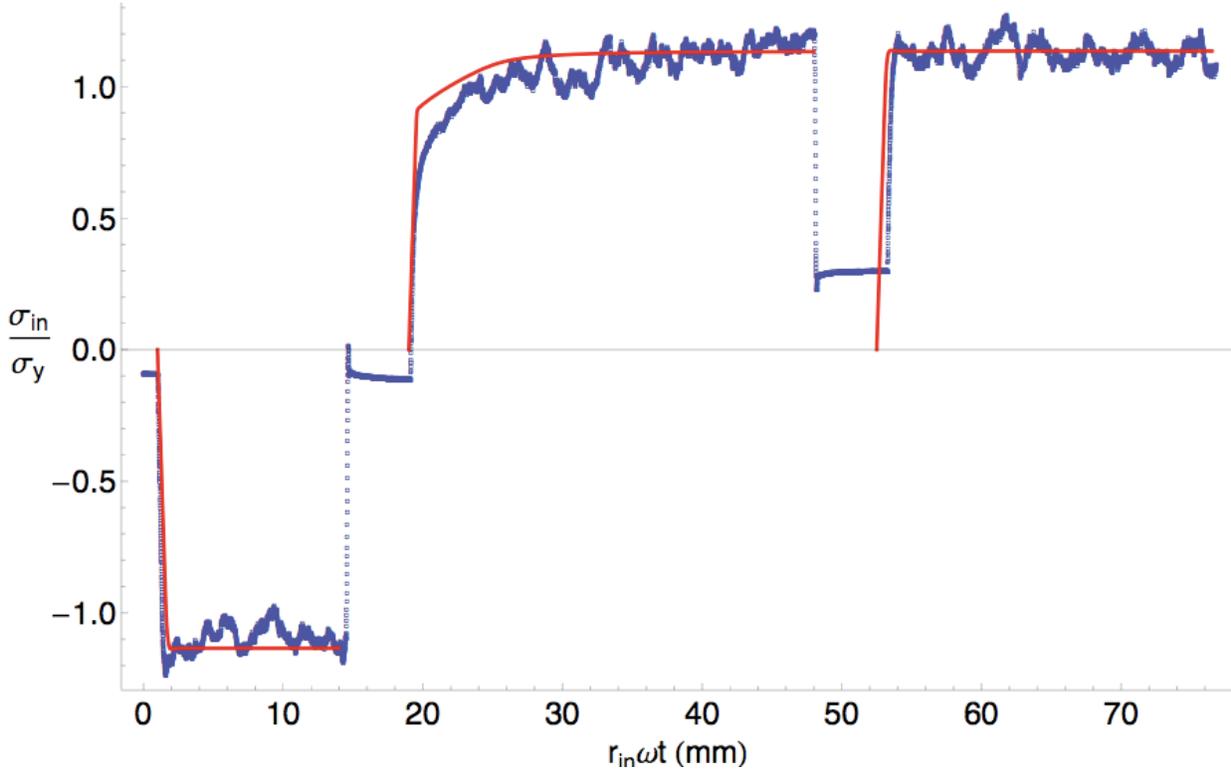

Figure 4

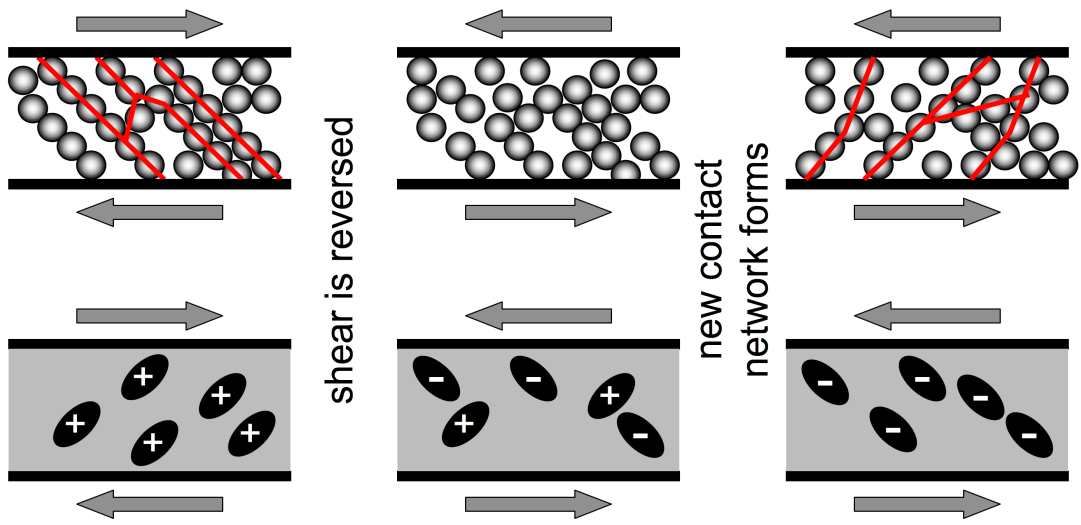

[17]  E. Bouchbinder, J. S. Langer, and I. Procaccia, *Phys. Rev. E* **75**, 036107 (2007).

[18]  M. L. Manning, J. S. Langer, and J. M. Carlson, *Phys. Rev. E* **76**, 056106 (2007).

[19]  Y. F. Shi,  M. B. Katz, H. Li, and M. L. Falk, *Phys. Rev. Lett.* **98**, 185505 (2007).

[20]  B. Utter, and R. P. Behringer, *European Phys. J. E* **14**, 373-380 (2004).

[21]  M. L. Manning, E.G. Daub, J. S. Langer and J. M. Carlson, Phys. Rev. E **79**, 016110 (2009).

[22]  T. S. Majmudar, and R. P. Behringer, *Nature* **435**, 1079-1082 (2005).

[23] M. Lundberg, K. Krishan, N. Xu, C. S. O'Hearn and M. Dennin, *Phys. Rev. E* 77, 041505 (2008).21